\documentclass[british]{article}
\usepackage[T1]{fontenc}
\usepackage[latin9]{inputenc}
\usepackage{geometry}
\usepackage{textcomp}
\usepackage{amsbsy}
\usepackage{graphicx}
\usepackage{setspace}
\usepackage{babel}
\begin{document}
\title{How do longitudinal waves propagate transversely?}
\author{P Huthwaite}

\maketitle
--- 

Acknowledgement: I prepared this paper in early 2021, and sent it
to a journal for review. The reviewers correctly identified that this
was, in fact, not an undiscovered phenomenon but that it was already
known by areas of the community. In particular, a reference was given
to {[}Catheline \& Benech, \textquotedbl Longitudinal shear wave
and transverse dilatational wave in solids\textquotedbl , JASA 137,
EL200-EL205 (2015){]}, and there were references within that dating
back to the 1990s. The other reviewer stated that this had been known
since 1956, and did not warrant a new publication. Any PhD students
reading this can maybe learn about the value of good literature reviews!
(Although finding appropriate keywords to search for in this case
is likely to be a challenge.)

I am posting this paper here, with this acknowledgement that I am
not claiming novelty, because I think it is an interesting and surprising
result, which does not seem to be widely known (certainly it was new
to anyone I spoke to about the problem). This phenomenon does appear
and causes some confusion when using finite element methods. 

---
\begin{abstract}
Modern, high-fidelity numerical simulations have shown an apparently
anomalous result: a longitudinal elastodynamic wave travelling perpendicular
to the forcing direction. Numerical simulations, in combination with
an analytical model, are used to confirm that this is not simply a
simulation artefact, but a true physical phenomenon, as well as illustrating
that the behaviour also occurs for shear and guided waves. When assessing
how this unexpected wave interacts with objects, however, it is found
that it does not scatter, despite being clearly measurable. This paper
has uncovered and explained this behaviour, which is critical for
the reliable use of numerical wave simulations in elastic media, as
well as potentially for high sensitivity experiments and other polarised
wave modalities.
\end{abstract}

\section{Introduction}

A longitudinal wave excited in an elastic medium is defined as travelling
in the same direction as the displacement within the wave. However,
recent high-accuracy simulations have highlighted an apparent contradiction
of this definition, showing that there is a small but measurable wave
travelling at the longitudinal wave speed completely perpendicularly
to the displacement direction, where only a shear wave (at different
speed) would be expected to be present. Similarly, it is possible
to identify a wave propagating at shear wave speed in the longitudinal
direction.

Guided wave simulations demonstrate the same behaviour. An in-plane
force excitation on the surface of the plate below the first cutoff
frequency would be expected to excite the S0 mode propagating along
the direction of excitation (and a small amount of A0), and SH0 perpendicular
to this, however, in both directions there appears to be a mixture
of both S0 and SH0.

The identification of these waves has only been possible through the
use of high-fidelity numerical methods for ultrasonic simulation in
solids, in particularly the finite element (FE) method, which has
grown hugely over the previous decades, both in the size of domains
modelled and the accuracy of the results. A consequence is that these
have uncovered new, unexpected behaviour which needs to be understood,
of which the waves discussed above form an important example. The
early cases of the use of FE in the late 1980s and early 1990s were
relatively isolated and were necessarily small given hardware constraints
(e.g.~\cite{Ludwig1988,Alleyne1992a}), then in the 2000s, as computational
power increased, the potential of simulations was exploited for developments
in ultrasound in areas such as air-coupled systems, guided wave approaches
and visco-elastic systems \cite{Castaings2004,Rajagopal2008,Ke2009,Zhang2011}.
Through the development of new techniques, including graphics-card-based
simulation tools \cite{Huthwaite:2014}, numerical methods in recent
years have delivered important advances in areas requiring high-fidelity
wavefield modelling, including Monte Carlo studies of scattering from
rough cracks \cite{Haslinger2020}, geometrical inversion techniques
\cite{Shi2019}, 3D simulations of guided waves in composites \cite{Leckey2014},
3D simulations of feature-guided waves \cite{Yu2019}, high-order
guided wave mode thickness measurement \cite{Belanger2014}, super-resolution
imaging \cite{Elliott2019} and high accuracy guided-wave tomographic
inversions \cite{Huthwaite2016a,Zimmermann2021}. Of particular note
is that multi-billion degree-of-freedom 3D models for grain scattering
have been run, minimising errors to the order of 0.01\% in wave speed
(see Fig.~7 of \cite{Huang2020}). Understanding the different waves
which exist in such models is critical to the success of mechanical
wave simulation across different applications areas including non-destructive
evaluation, seismology and medicine. 

Such high-fidelity simulations have highlighted the presence of the
wave which is the focus of this paper: the longitudinal wave apparently
travelling transversely. To the author's knowledge, the presence of
this wave has not been previously recognised, which leaves an important
question about whether it is a genuine physical phenomenon, revealed
by the increasing accuracy of modern numerical models, or simply a
simulation artefact. This paper will use numerical simulations to
demonstrate the phenomenon and highlight its important features, but
also provide analytical calculations to confirm it is genuine. This
knowledge is vital for numerical models (potentially as well as very
accurate experiments) because the presence of an unexpected wave mode
could indicate a fundamental error in the excitation or measurement
approach which needs to be addressed, possibly at significant cost,
so understanding that this occurs and when it does so is important.
The consequences of the wave's existence must also be established:
what impact does it practically have when it begins interacting with
other objects?

This paper will firstly demonstrate the presence of the wave with
the FE method in Sect.~\ref{sec:FEdemo}, and an analytical model
will explain the origin of the wave in Sect.~\ref{sec:Analytical-model}.
Section \ref{sec:Scattering-of-wave} will analyse the scattering
behaviour of this wave, then Sect.~\ref{sec:Guided-wave-demonstration}
will evaluate the behaviour of sources for guided waves.

\section{Demonstration of the phenomenon via numerical modelling \label{sec:FEdemo}}

\subsection{Model setup}

\begin{figure}
\includegraphics[width=0.7\textwidth]{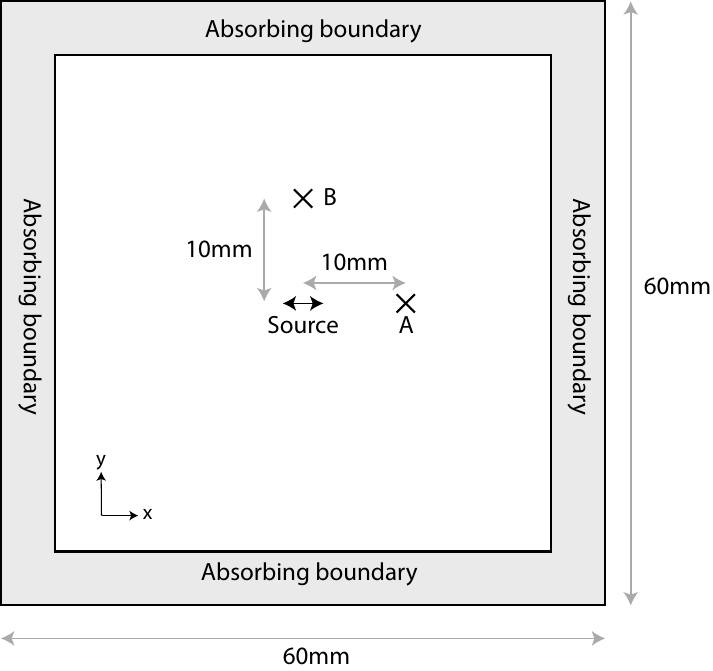}

\caption{\label{fig:model2d}Initial 2D model for demonstrating the presence
of the perpendicularly travelling longitudinal wave.}

\end{figure}
 Numerical modelling is used to demonstrate the behaviour of the waves
radiating from a point source. The physical setup, illustrated in
Fig.~\ref{fig:model2d}, was taken to be a 2D plane strain elastic
medium, with a point source at its centre: a force acting in the $x$
direction, which was excited with 3 cycles (Hann windowed) of a 1.5MHz
sinusoidal excitation. The material was set to have properties of
steel, with Young's modulus 210GPa, shear modulus 80GPa and density
8000kg/m\textsuperscript{3}; at the centre frequency the longitudinal
and shear wavelengths become 4mm and 1.6mm respectively. Time trace
measurements were taken at two locations: A which is 10mm across in
$x$, and B which is offset 10mm in the $y$ direction.

A finite element elastodynamic simulation was used to simulate the
wavefield, with a mesh of element size 0.05mm in both dimensions,
and 1200 4-noded linear quad elements in each direction giving a grid
of 60$\times$60mm\textsuperscript{2}; this element size gives 81
elements per wavelength for the longitudinal waves and 33 for the
shear wavelength. These elements had reduced integration with hourglass
control. A time step of 6.6ns was selected, giving a Courant number
of 0.8, and absorbing boundaries using the Stiffness Reduction Method
\cite{Pettit2014} were applied to all four boundaries over a 12mm
(3$\lambda$ at the centre frequency for longitudinal waves) distance.
The graphics-card-based explicit time domain solver Pogo was used
to solve this \cite{Huthwaite:2014}; on an Nvidia GeForce GTX 1080
Ti (a consumer card aimed at the gaming market) the 1.44 million-node
model with 1362 time increments ran in 3.1 seconds. 

\subsection{Results}

\begin{figure}
\includegraphics[width=1\textwidth]{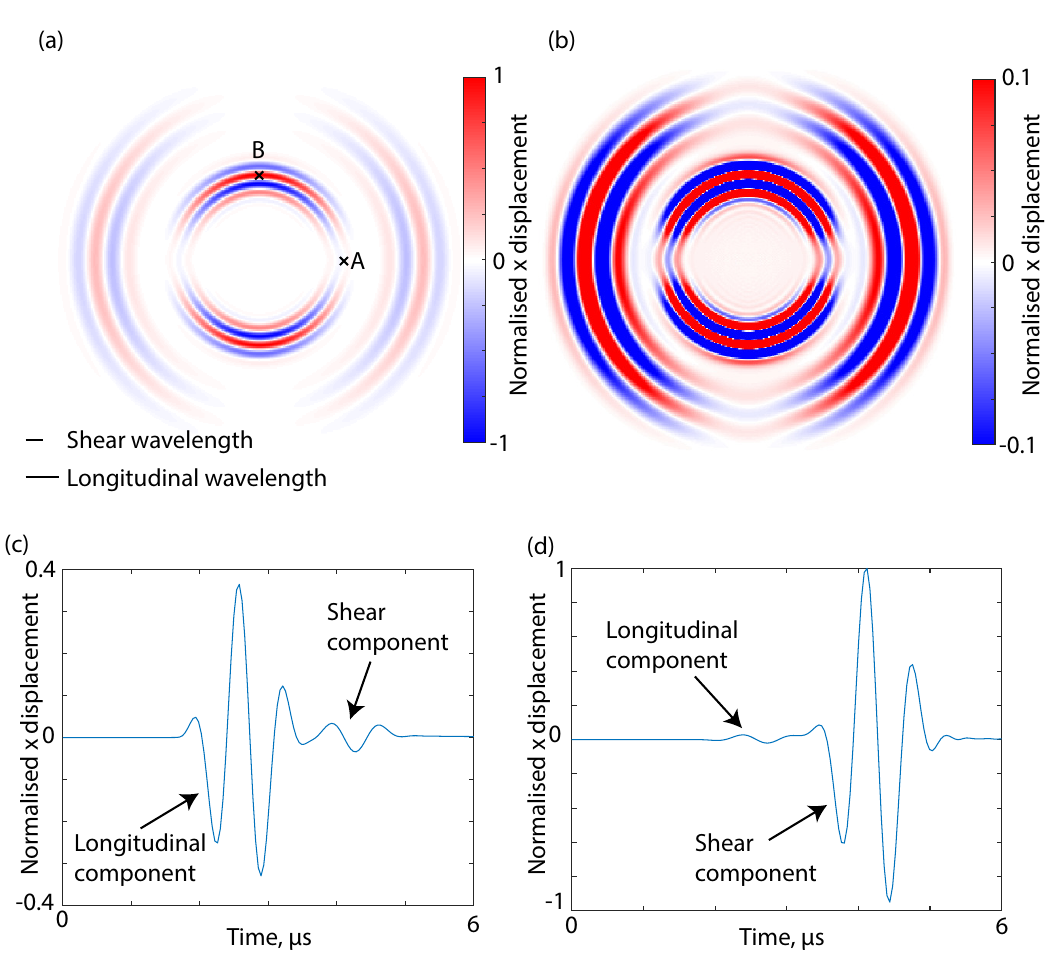}

\caption{\label{fig:FEField}$x$ displacement values of the finite element
simulation of a point source, with displacement values scaled relative
to the peak displacement at 4$\mu$s. (a) shows the field at 4$\mu$s
colours scaled across the full range while (b) shows the scale adjusted
to 10\% such that the presence of the longitudinal and shear components
in all directions are visible. (c) and (d) plot the x direction time
traces measured at points A and B marked in (a) respectively.}

\end{figure}

Figure \ref{fig:FEField} shows the $x$ component of the displacement
field at 4$\mu$s; Fig.~\ref{fig:FEField}(a) with the colours scaled
to capture the full range and Fig.~\ref{fig:FEField}(b) with the
scale adjusted to significantly clip the peak values and show the
presence of both the longitudinal and shear waves at all angles, highlighting
the key result of this paper. There are some notable features in these
plots. The two wave modes appear to exhibit the same behaviour, with
a 90 degree rotation. There is a clear phase advancement of the longitudinal
waves along the y axis compared to the waves in the other directions,
and this is visible in the shear waves too. The amplitude does drop
off significantly in the direction perpendicular to the propagation
of longitudinal waves, where one would expect to see no waves of this
mode, although as illustrated it does not reach zero, and the same
behaviour is true for shear. Figures \ref{fig:FEField}(c) and (d)
illustrate the time traces of the $x$ displacement taken at positions
A and B from Fig.~\ref{fig:FEField}(a) respectively, confirming
the results of the field plots; in both perpendicular directions,
both modes exist and are clearly measurable. The next section will
attempt to mathematically illustrate and explain this.

\section{Analytical model\label{sec:Analytical-model}}

Consider a point excitation in the $x$ direction in a 2D elastic
domain. Physically, this will correspond to a force in the $x$ direction,
matching the FE simulation in Sect.~\ref{sec:FEdemo}. Taking the
scalar potential of the longitudinal wave, this source can be expressed
as a dipole of the form
\begin{equation}
s_{L}\left(\boldsymbol{x}\right)=A_{L}\frac{\delta\left(\boldsymbol{x}-\boldsymbol{x}_{0}-\epsilon\boldsymbol{i}\right)-\delta\left(\boldsymbol{x}-\boldsymbol{x}_{0}+\epsilon\boldsymbol{i}\right)}{k\epsilon}\label{eq:longSource}
\end{equation}
assumed centred at $\boldsymbol{x}_{0}$ and with a small spacing
of $2\epsilon$ between the two component point sources. $k$ is the
wavenumber, $\boldsymbol{i}$, $\boldsymbol{j}$ will be used as standard
unit vectors in the $x$ and $y$ directions respectively, and $A_{L}$
describes the strength of this source. Similarly, the shear potential
source will be described as
\begin{equation}
s_{S}\left(\boldsymbol{x}\right)=A_{S}\frac{\delta\left(\boldsymbol{x}-\boldsymbol{x}_{0}-\epsilon\boldsymbol{j}\right)-\delta\left(\boldsymbol{x}-\boldsymbol{x}_{0}+\epsilon\boldsymbol{j}\right)}{k_{sh}\epsilon}
\end{equation}
with $A_{S}$ being the shear wave source strength. For now just the
longitudinal component will be considered, which will be independent
of the shear. The field radiated from a single unit point source is
the Green's function, which in 2D free space becomes
\begin{equation}
G\left(\boldsymbol{x}-\boldsymbol{x}_{0}\right)=H_{0}^{(1)}\left(k\left|\boldsymbol{x}-\boldsymbol{x}_{0}\right|\right)
\end{equation}
or taking the radius $r=\left|\boldsymbol{x}-\boldsymbol{x}_{0}\right|$
as being larger than a wavelength then this can be approximated as
\begin{equation}
G_{\infty}\left(\boldsymbol{x}-\boldsymbol{x}_{0}\right)=\Pi\frac{e^{ikr}}{\sqrt{r}}
\end{equation}
with 
\[
\Pi=\frac{e^{-i\pi/4}}{\sqrt{8\pi k}}.
\]
Applying this to the dipole source, the constant $\Pi$ will be dropped,
parameters set as $A_{L}=1$, and $\boldsymbol{x}_{0}=\boldsymbol{0}$
and the two radii expressed as $r_{a}=\left|\boldsymbol{x}-\epsilon\boldsymbol{i}\right|$
and $r_{b}=\left|\boldsymbol{x}+\epsilon\boldsymbol{i}\right|$ to
give the scalar potential of the field as
\begin{equation}
\phi=\frac{e^{ikr_{a}}}{k\epsilon\sqrt{r_{a}}}-\frac{e^{ikr_{b}}}{k\epsilon\sqrt{r_{b}}}.
\end{equation}
Since the separation $\epsilon$ is small, the slight discrepancy
between radii in the denominator terms can be neglected, i.e.~let
$r\approx r_{a}\approx r_{b}$
\begin{equation}
\phi=\frac{e^{ikr_{a}}-e^{ikr_{b}}}{k\epsilon\sqrt{r}}
\end{equation}
then for the exponential parts the approximations $r_{a}\approx r-\epsilon\cos\theta$
and $r_{b}\approx r+\epsilon\cos\theta$ can be utilised, with $\theta$
being the angle of position $\boldsymbol{x}$ relative to the $x$
axis. Incorporating this,
\begin{equation}
\phi=e^{ikr}\frac{e^{ik\epsilon\cos\theta}-e^{-ik\epsilon\cos\theta}}{k\epsilon\sqrt{r}}
\end{equation}
which becomes
\begin{equation}
\phi=e^{ikr}\frac{2i\sin\left(k\epsilon\cos\theta\right)}{k\epsilon\sqrt{r}}.
\end{equation}
Then, since $k\epsilon$ is small, the approximation $\sin\alpha=\alpha$,
which is valid at the asymptote, can be used to simplify this to:
\begin{equation}
\phi=e^{ikr}\frac{2i\cos\theta}{\sqrt{r}}.\label{eq:potentialField}
\end{equation}
Now consider what would happen on the $y$ axis, where $\theta=\pi/2$.
At this point, $\cos\theta=0$ so the final result is $\phi_{L}=0$.
This result in the potential field clearly does drop down to zero,
which is the intuitive expectation. However, it is now possible to
investigate what happens to the displacement field itself at this
location. This is defined as 
\begin{equation}
\boldsymbol{u}=\boldsymbol{\nabla}\phi+\boldsymbol{\nabla}\times\boldsymbol{\psi}.
\end{equation}
The shear component, $\boldsymbol{\psi}$, will continue to be ignored
for now and just the $\boldsymbol{\nabla}\phi$ term considered. In
polar coordinates this becomes
\begin{eqnarray}
\boldsymbol{u_{L}}=\boldsymbol{\nabla}\phi & = & \frac{\partial\phi}{\partial r}\boldsymbol{e_{r}}+\frac{1}{r}\frac{\partial\phi}{\partial\theta}\boldsymbol{e_{\theta}}\\
 & = & 2i\cos\theta\left(\frac{ike^{ikr}}{\sqrt{r}}-\frac{e^{ikr}}{2r\sqrt{r}}\right)\boldsymbol{e_{r}}+\frac{1}{r}\frac{2i}{\sqrt{r}}e^{ikr}\left(-\sin\theta\right)\boldsymbol{e_{\theta}}\\
 & = & \frac{-2}{\sqrt{r}}e^{ikr}\left[\left(k+\frac{i}{2r}\right)\cos\theta\boldsymbol{e_{r}}+\frac{i}{r}\sin\theta\boldsymbol{e_{\theta}}\right]\label{eq:longUSoln}
\end{eqnarray}
Here $\boldsymbol{e_{r}}$ and $\boldsymbol{e_{\theta}}$ are unit
vectors in the radial and circumferential directions respectively.
It is clear from this that the $\boldsymbol{e_{r}}$ component, like
the scalar potential, disappears to zero at $\theta=\pi/2$ due to
the $\cos\theta$ dependence. However, the $\boldsymbol{e_{\theta}}$
term does not; along the $y$ axis the values become
\begin{equation}
\boldsymbol{u_{L}}\left(\theta=\pi/2\right)=-\frac{2ik}{r^{3/2}}.
\end{equation}
At this point $\boldsymbol{e_{\theta}}=\boldsymbol{i}$, i.e.~this
component is aligned with the $x$ axis so the wave oscillation will
be in this direction. The wave also drops off in amplitude by a factor
of $1/r$, in addition to the $1/\sqrt{r}$ which occurs with beam
spread, and the additional $i$ factor causes a phase shift of $+\pi/2$
relative to the main wave in the $\boldsymbol{e_{r}}$ direction,
dominated by the $k$ term. This matches the behaviour identified
in Sect.~\ref{sec:FEdemo}. Before establishing the consequences
of this, the analysis is completed by discussing the shear wave. Following
the logic for the longitudinal waves, the dipole field for the shear
source can be described as
\begin{equation}
\psi_{z}=e^{ik_{s}r}\frac{2i\sin\theta}{\sqrt{r}}
\end{equation}
where $k_{s}$ is the shear wavenumber, and just the out-of-plane,
$z$, component $\psi_{z}$ is considered to be nonzero for this 2D
problem. Utilising the derivation of curl in polar coordinates:
\begin{eqnarray}
\boldsymbol{u_{S}}=\boldsymbol{\nabla}\times\boldsymbol{\psi} & = & \frac{1}{r}\frac{\partial\psi_{z}}{\partial\theta}\boldsymbol{e_{r}}-\frac{\partial\psi_{z}}{\partial r}\boldsymbol{e_{\theta}}\\
 & = & \frac{1}{r}\frac{2i}{\sqrt{r}}e^{ik_{s}r}\cos\theta\boldsymbol{e_{r}}-2i\sin\theta\left(\frac{ik_{s}e^{ik_{s}r}}{\sqrt{r}}-\frac{e^{ik_{s}r}}{2r\sqrt{r}}\right)\boldsymbol{e_{\theta}}\\
 & = & \frac{2}{\sqrt{r}}e^{ik_{s}r}\left[\frac{i}{r}\cos\theta\boldsymbol{e_{r}}+\left(k_{s}+\frac{i}{2r}\right)\sin\theta\boldsymbol{e_{\theta}}\right].\label{eq:shearUSoln}
\end{eqnarray}
Perhaps unsurprisingly, the equation is clearly very similar to eq.~(\ref{eq:longUSoln})
derived for the longitudinal waves; the differences are that the system
is rotated by $\pi/2$ and the shear wavenumber is used rather than
longitudinal. 

\begin{figure}
\includegraphics[width=1\textwidth]{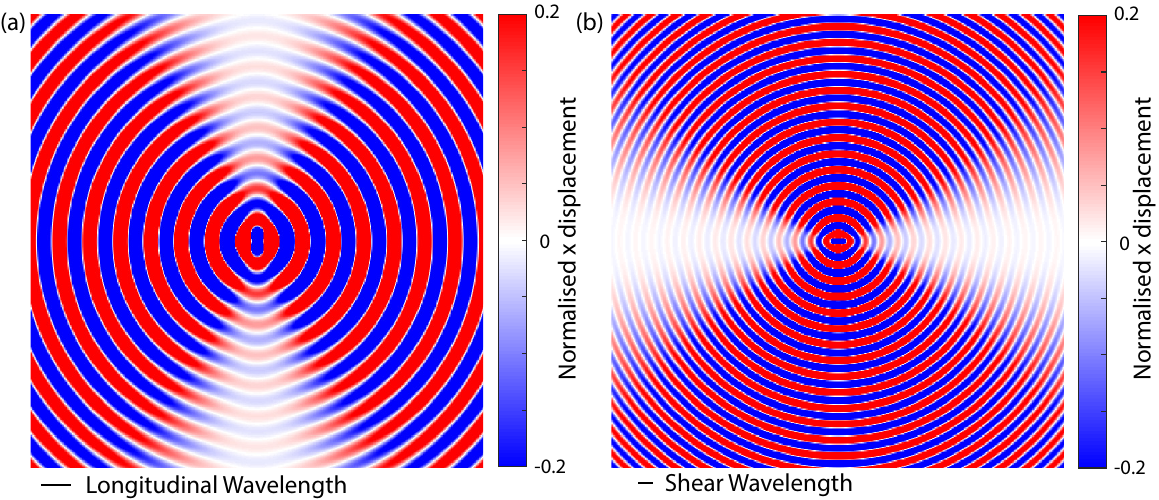}

\caption{\label{fig:analyticalFields}Analytical wavefield solutions for (a)
longitudinal and (b) shear waves, given by eqs.~(\ref{eq:longUSoln})
and (\ref{eq:shearUSoln}) respectively, calculated in steel (properties
given in Sect.~\ref{sec:FEdemo}) at 1.5MHz. Colours for both wavefields
are clipped to 20\% of the peak amplitude measured at 5 wavelengths
from the source.}

\end{figure}

Figure \ref{fig:analyticalFields} plots the calculated fields for
the longitudinal (Fig.~\ref{fig:analyticalFields}(a)) and shear
(Fig.~\ref{fig:analyticalFields}(b)) components. Both illustrate
similar behaviour to that visible in the equivalent FE solution, Fig.~\ref{fig:FEField},
although it is noted that the FE solution is a single frame of a time
domain simulation while the analytical version is a single frequency
solution, i.e.~is solved for a continuous wave. As before, there
is a clear advance in the wave propagating perpendicular to the main
direction of travel (will be referred to as the ``perpendicular wave''
for the rest of this discussion), reflecting the additional $i$ term
causing a phase shift in this component, confirming the FE behaviour.

There are some interesting observations which can be made. On one
level this could be considered a near field effect; the relative drop
off of $1/kr$ means that its effect will halve as distance from the
source is doubled, so at very large distances it will not be significant.
However, such identification of ``near'' and ``far'' fields are
usually reserved for distinct regions of behaviour, which is not the
case here, so this is less appropriate.

If there was an infinite line source in the $x$ direction then the
derivative of the longitudinal potential field in $x$ must become
zero, i.e.~$\partial\phi/\partial x=0$, since there can be no variation
in this direction. This means that the corresponding displacement
must drop to zero, and therefore the perpendicular wave will not exist.
As a result, the perpendicular wave must occur because the wavefield
is not straight, i.e.~it is the curvature of the wavefield which
causes the derivative to be nonzero and hence produces the measurable
component. The presence of the $1/r$ factor is therefore unsurprising;
it indicates that the strength of the component is inversely proportional
to the radius of curvature. Therefore, it is possible to also predict
that for any wave of this form, any curvature will result in a visible
wave appearing as the derivative becomes nonzero.

It is possible to hypothesise physical justifications for the presence
of the component. It could be considered a diffractive effect, where
the wavefronts curve around to join up with each other. The two halves
of the domain have symmetrical wavefields and diffraction conspires
to, in effect, fill the join between them. It should be recognised
that for this to be attributed to diffraction, the effect should not
be present in the geometrical regime, i.e.~at high frequencies. In
both eqs.~(\ref{eq:longUSoln}) and (\ref{eq:shearUSoln}) at large
frequencies, $k$ and $k_{sh}$ will be large, and thus dominate the
equation, leaving the perpendicular wave component to vanish to zero
in the geometric limit. This acts as a mathematical confirmation that
the phenomenon can be considered a form of diffraction. 

\section{Scattering of wave\label{sec:Scattering-of-wave}}

Previously it has been seen that the longitudinal wave is measurable
at a location perpendicular to the source direction; this is the case
both in numerical simulations and the equivalent analytical solution.
However, this does raise a question about what the consequences are
for the behaviour of the wave, i.e.~while it can clearly be measured
at this location, what, if any, effect will it have when it interacts
with an object?

\begin{figure}
\includegraphics[width=0.8\textwidth]{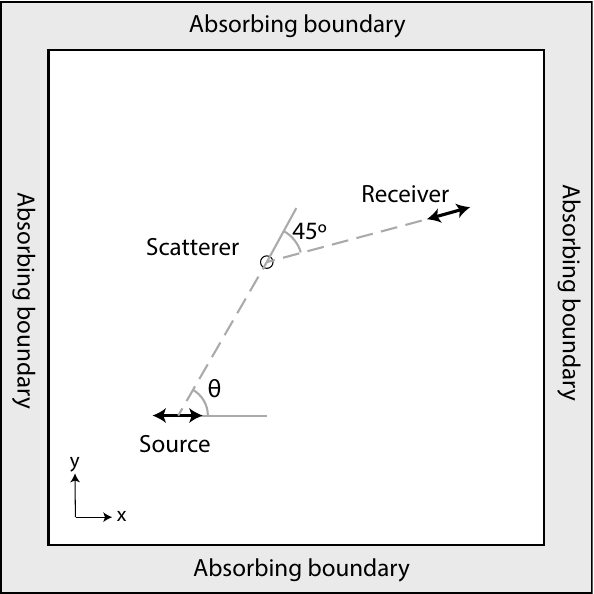}

\caption{\label{fig:scatteringModel}Schematic of model to test scattering
behaviour in all directions. The scatterer and receiver are rotated
rigidly around the source.}

\end{figure}

To evaluate this, a set of test simulations are introduced, to evaluate
how the wave will behave as it meets a point scatterer located at
different angular locations. The simulation is set up as illustrated
in Fig.~\ref{fig:scatteringModel}. A model of size 100$\times$100mm\textsuperscript{2}
was defined, with a 1.5MHz source positioned 30mm in each dimension
from the lower left corner. A small scatterer was then placed a further
30mm from this source at an angle $\theta$ which could be adjusted
around the source, to enable all angles to be sampled. To form this
scatterer, a hollow circle of size 0.2mm was used, which is small
enough to be treated as a point\footnote{The author has also demonstrated identical behaviour with a scatterer
formed by fixing both $x$ and $y$ displacements but these results
are not presented here.}. An 8mm absorbing boundary using the SRM \cite{Pettit2014} was applied
around the edges of the domain.

The strength of the resulting scattered field should then be measured.
It is noted that the scattered field will vary as a function of the
scattering angle relative to the incident wave, so it is important
to keep this constant, as well as maintaining the same distance from
the scatterer. However, since it is not of interest to obtain absolute
values of scattering amplitude, but rather how this changes as the
scatterer moves around the source, the specific angle at which the
scattered field is measured relative to the source is unimportant,
provided it is constant. For these purposes, a scattering angle of
45º was used throughout, as shown in Fig.~\ref{fig:scatteringModel},
with the scatterer and its measurement point being rigidly rotated
around the source. The 45º angle was selected as this should produce
both longitudinal and shear waves and hence avoid any scenario where
the receiver was positioned to miss one particular type of wave. The
receiver was set to measure displacement in the direction directly
aligned with the scatterer as shown in the figure.

Two sets of simulations are run. The first set was as discussed above,
and the second was an incident case where the circle was filled but
the remainder of the mesh was unchanged, effectively removing the
scatterer. The difference between these two cases is taken to give
the scattered field, which is the final quantity considered.

A free mesh was used to model the domain, allowing arbitrary positioning
of the source, scatterer and receivers. The target element size was
set to 0.04mm, to give 100 elements per longitudinal wavelength. The
elements were linear triangular plane strain elastic elements. Both
the total and incident models were run 101 times each to sweep through
all angles of $\theta$ from 0° to 100° at 1° intervals. Once loaded,
the incident component was subtracted and the signals were windowed
to obtain just the first arrivals, corresponding to the longitudinal
excited waves scattered to longitudinal at the scatterer. These were
Fourier transformed and the component corresponding to 1.5MHz extracted,
normalised by dividing by their mean across the 101 values, and plotted. 

\begin{figure}
\includegraphics[width=1\textwidth]{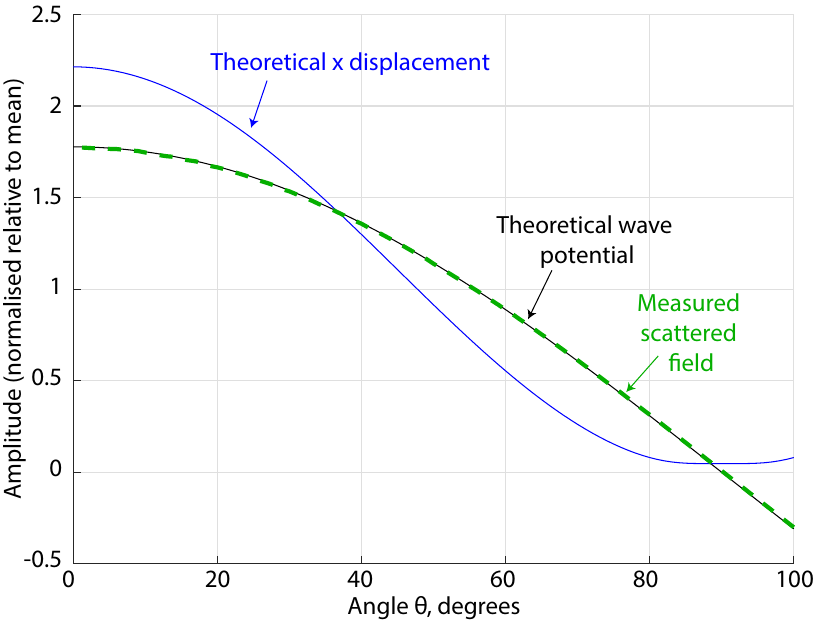}

\caption{\label{fig:scatterTest}The measured scattered field from the setup
of Fig.~\ref{fig:scatteringModel} compared to the theoretically
calculated values for wave potential from eq.~(\ref{eq:potentialField})
and displacement from resolving eq.~(\ref{eq:longUSoln}) into the
$x$ direction.}

\end{figure}

Figure \ref{fig:scatterTest} shows a comparison between the scattered
field measured and the different wave components calculated analytically
in Sect.~\ref{sec:Analytical-model}. These demonstrate how the scattered
field is clearly following the pattern of the wavefield potential,
rather than the $x$ component of the wave, notably passing through
zero amplitude at the 90º point. This indicates that the wave propagating
in the perpendicular direction does not scatter, despite existing
as a displacement. This leads to an interesting, somewhat philosophical,
conclusion: while the wave does exist at that location, in that it
is clearly measurable, it does not exist in that it does not cause
any scattering. 

Hypothesising about the reason for this behaviour, it can be stated
that the underlying scalar potential is the fundamental quantity,
and the physical displacement fields produced arise as a result of
this potential field. Based on this, it would seem logical that scattering
would be proportional to this scalar potential rather than any other
derived quantity, even though the scalar potential is not directly
measurable itself. 

\section{Guided wave demonstration\label{sec:Guided-wave-demonstration}}

Guided waves have important applications within NDE, having some desirable
properties including their ability to propagate large distances with
minimal amplitude loss, enabling large components to be inspected.
The phenomenon described in this paper also exists for guided waves
in plates for the S0 and SH0 waves; physically this is unsurprising
since an S0/SH0 model can be captured well in the low frequency regime
by 2D plane stress behaviour, and the only practical difference between
this and plane strain considered for bulk waves is that the effective
stiffnesses change. This section will provide a confirmation of this
behaviour through numerical modelling, as well as including analysis
of the A0 mode.

\begin{figure}
\includegraphics[width=1\textwidth]{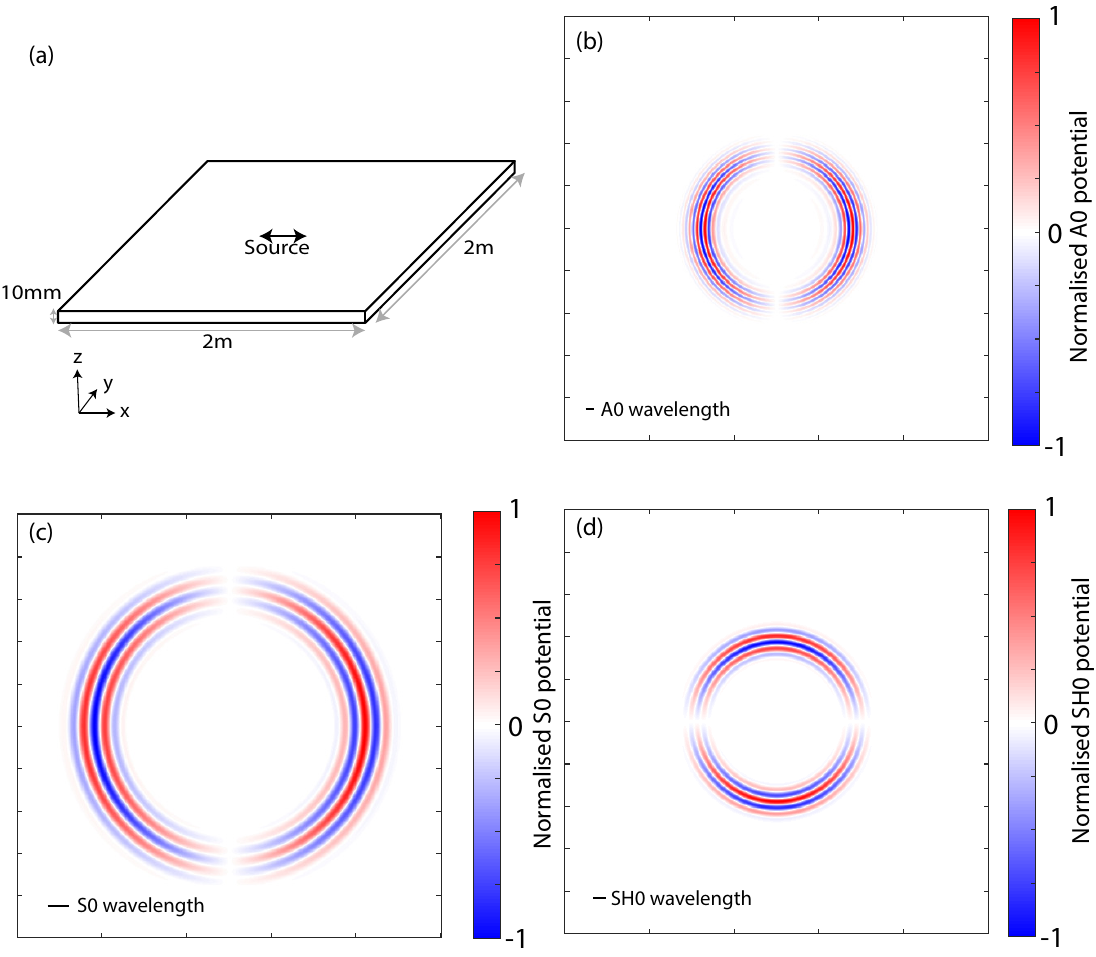}

\caption{\label{fig:gwScalars}(a) Illustration of the physical guided wave
model. Guided wave scalar potentials for (b) A0, (c) S0, (d) SH0,
normalised relative to their respective peaks. These are taken at
a time of 151$\mu$s.}

\end{figure}

A finite element model was again set up, a 3D plate of 10mm thickness
($z$ direction) and extent 2m in the in-plane directions, $x$ and
$y$, as illustrated in Fig.~\ref{fig:gwScalars}(a). A 1N force
was excited in the $x$ direction, centred in the the $x$ and $y$
directions and applied to the top surface of the plate; this was a
Hann-windowed toneburst of 3 cycles and a centre frequency of 50kHz.
This excitation will excite all three of the fundamental modes, A0,
S0 and SH0. The element size was set to 2mm in each dimension giving
over 60 8-node linear brick elements per longitudinal bulk wavelength
and 31 for shear, which should be more than sufficient for the resulting
guided waves, and the time step was 0.165$\mu$s targetting a Courant
number of 0.5. No absorbing boundaries were used in this model.

Figures \ref{fig:gwScalars}(b), (c) and (d) show the scalar potentials
for the three different modes present, A0, SH0 and S0, calculated
from the output displacement fields in $x$, $y$ and $z$. The A0
mode was extracted by taking the difference in the divergence in $x$
and $y$ on the top and bottom surfaces, S0 was extracted from the
difference between the $z$ components on top and bottom surfaces,
and SH0 was taken as the curl in $x$ and $y$ at the central plane
of the plate. In all cases the necessary differentials were calculated
as simple first-order finite differences between adjacent nodes on
the grid. In each case the behaviour is as expected; the potential
disappears to zero (and the sign changes) perpendicular to the direction
of propagation.

\begin{figure}
\includegraphics[width=1\textwidth]{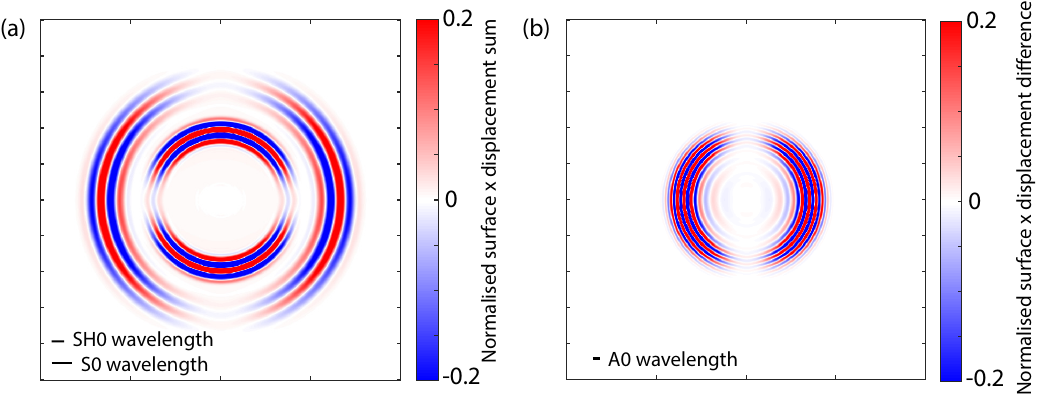}

\caption{\label{fig:displacementsGuidedWaves}$x$ displacement values (a)
summed on the top and bottom surface to show SH0 and S0, (b) bottom
surface subtracted from top to show A0. Both plots are normalised
against their respective peaks, and the colours are then clipped to
20\% of the maximum. }
\end{figure}

Figure \ref{fig:displacementsGuidedWaves} then illustrates the displacements
which correspond to the scalar potentials. While plotting the top
$x$ displacement would be sufficient to visualise the waves, the
A0 and SH0 waves have similar velocities at this frequency-thickness
and would overlap. To address this, the asymmetry of A0 versus the
symmetry of SH0 have been exploited to separate: the sum of displacement
on top and bottom is illustrated in Fig.~\ref{fig:displacementsGuidedWaves}(a)
capturing the symmetry of S0 and SH0, and the displacement difference
is shown in Fig~\ref{fig:displacementsGuidedWaves}(b) capturing
just A0. 

In all three cases the colour scale has been adjusted so that the
presence of the perpendicularly travelling component is clearly visible.
As discussed previously, this is to be expected for the S0 and SH0
modes via the plane strain/plane stress argument. However, it appears
for the bending mode A0 too. 

As a generalisation of the principle, there are two key aspects which
can be identified from the derivation of Sect.~\ref{sec:Analytical-model}
as being necessary for the presence of the ``perpendicular wave''
to exist:
\begin{enumerate}
\item The source must form a dipole within the scalar formulation for that
specific wave.
\item The physical wavefield must be some form of derivative of the underlying
scalar potential.
\end{enumerate}
The first rule provides the zero in the direction perpendicular to
the primary direction of travel, but the derivative itself will not
be zero, such that the second rule enables a wave to be physically
measured. The dipolar source implies that the wave must be polarised.
This can be either a transverse wave or longitudinal, i.e.~direction
of motion is the same as the direction of propagation; through this
paper both forms have been demonstrated, as well as the A0 mode which
is a combination of both forms. 

\section{Conclusions}

This paper has identified and explained a counter-intuitive wave component
which propagates perpendicularly to the expected direction of propagation,
and highlighted its existence for both longitudinal and shear waves,
as well as for guided waves in plates. The underlying wavefield potentials,
arising from the Helmholtz decomposition, do reduce to zero in the
perpendicular direction as expected, but since their derivatives do
not, the physical displacement itself does not. The waves have interesting
properties: while they are demonstrably measurable, they do not scatter,
and it has been shown that they are caused through a diffractive effect.
Thoroughly uncovering the phenomenon is critical for the reliable
use of high-fidelity numerical simulations.

This wave phenomenon is also likely to occur with other physical modalities
which are polarised, provided the source forms a dipole in the scalar
potential and that the physical wavefield is then a derivative of
this. This should be the case with electromagnetic waves, although
such investigations are beyond the scope of this paper.

\section{Acknowledgements}

The author would like to thank Mike Lowe, Peter Cawley, Stefano Mariani
and Euan Rodgers variously for raising the issue initially, discussions
around the phenomenon and comments on the draft of this paper. 

\section{Data availability}

The scripts to generate the finite element models and recreate the
results from this work are available from {[}Supplementary material{]}.

\bibliographystyle{ieeetr}
\bibliography{ph}

\end{document}